\documentclass[12pt,a4paper]{article}

\usepackage{ifthen} 
\newboolean{pdflatex}
\setboolean{pdflatex}{true} 

\usepackage{color}
\usepackage{rotating}
\usepackage{graphpap}
\usepackage{multirow}
\usepackage{afterpage}
\usepackage[normalem]{ulem}
\usepackage{mathrsfs}

\newboolean{articletitles}
\setboolean{articletitles}{true} 

\newboolean{uprightparticles}
\setboolean{uprightparticles}{true} 

\newboolean{inbibliography}
\setboolean{inbibliography}{false} 

\usepackage{microtype}
\usepackage{lineno}  
\usepackage{xspace} 
\usepackage{caption} 

\usepackage{graphicx}  
\usepackage{color}
\usepackage{colortbl}
\graphicspath{{./figs/}} 

\usepackage{amsmath} 
\usepackage{amssymb}
\usepackage{amsfonts}
\usepackage{upgreek} 

\newcommand*\patchAmsMathEnvironmentForLineno[1]{%
\expandafter\let\csname old#1\expandafter\endcsname\csname #1\endcsname
\expandafter\let\csname oldend#1\expandafter\endcsname\csname
end#1\endcsname
 \renewenvironment{#1}%
   {\linenomath\csname old#1\endcsname}%
   {\csname oldend#1\endcsname\endlinenomath}%
}
\newcommand*\patchBothAmsMathEnvironmentsForLineno[1]{%
  \patchAmsMathEnvironmentForLineno{#1}%
  \patchAmsMathEnvironmentForLineno{#1*}%
}
\AtBeginDocument{%
\patchBothAmsMathEnvironmentsForLineno{equation}%
\patchBothAmsMathEnvironmentsForLineno{align}%
\patchBothAmsMathEnvironmentsForLineno{flalign}%
\patchBothAmsMathEnvironmentsForLineno{alignat}%
\patchBothAmsMathEnvironmentsForLineno{gather}%
\patchBothAmsMathEnvironmentsForLineno{multline}%
}

\usepackage{hyperref}    
\usepackage[all]{hypcap} 

\def\lhcb {\mbox{LHCb}\xspace}

\ifthenelse{\boolean{uprightparticles}}%
{

 \def\Pvarepsilon {\ensuremath{\upvarepsilon}\xspace}                 
                  
 \def\Peta        {\ensuremath{\upeta}\xspace}

 \def\Pmu         {\ensuremath{\upmu}\xspace}                 
 \def\Pnu         {\ensuremath{\upnu}\xspace}                 
                  
 \def\Ppi         {\ensuremath{\uppi}\xspace}

 \def\Ptau        {\ensuremath{\uptau}\xspace}

 \def\Ppsi        {\ensuremath{\uppsi}\xspace}

 \def\PDelta      {\ensuremath{\Delta}\xspace}                 
 \def\PXi      {\ensuremath{\Xi}\xspace}                 
 \def\PLambda      {\ensuremath{\Lambda}\xspace}                 
 \def\PSigma      {\ensuremath{\Sigma}\xspace}                 
 \def\POmega      {\ensuremath{\Omega}\xspace}                 
 \def\PUpsilon      {\ensuremath{\Upsilon}\xspace}                 
 
 \def\PB      {\ensuremath{\mathrm{B}}\xspace}                 
                  
 \def\PD      {\ensuremath{\mathrm{D}}\xspace}

 \def\PJ      {\ensuremath{\mathrm{J}}\xspace}                 
 \def\PK      {\ensuremath{\mathrm{K}}\xspace}

 \def\PW      {\ensuremath{\mathrm{W}}\xspace}

 \def\Pb      {\ensuremath{\mathrm{b}}\xspace}                 
 \def\Pc      {\ensuremath{\mathrm{c}}\xspace}

 \def\Pi      {\ensuremath{\mathrm{i}}\xspace}

 \def\Pp      {\ensuremath{\mathrm{p}}\xspace}

 \def\Ps      {\ensuremath{\mathrm{s}}\xspace}

}
{

 \def\Pvarepsilon {\ensuremath{\varepsilon}\xspace}                 
                  
 \def\Peta        {\ensuremath{\eta}\xspace}

 \def\Pmu         {\ensuremath{\mu}\xspace}                 
 \def\Pnu         {\ensuremath{\nu}\xspace}                 
                  
 \def\Ppi         {\ensuremath{\pi}\xspace}

 \def\Ptau        {\ensuremath{\tau}\xspace}

 \def\Ppsi        {\ensuremath{\psi}\xspace}                 
                  
 \mathchardef\PDelta="7101
 \mathchardef\PXi="7104
 \mathchardef\PLambda="7103
 \mathchardef\PSigma="7106
 \mathchardef\POmega="710A
 \mathchardef\PUpsilon="7107
                  
 \def\PB      {\ensuremath{B}\xspace}                 
                  
 \def\PD      {\ensuremath{D}\xspace}

 \def\PJ      {\ensuremath{J}\xspace}                 
 \def\PK      {\ensuremath{K}\xspace}

 \def\PW      {\ensuremath{W}\xspace}

 \def\Pb      {\ensuremath{b}\xspace}                 
 \def\Pc      {\ensuremath{c}\xspace}

 \def\Pi      {\ensuremath{i}\xspace}

 \def\Pp      {\ensuremath{p}\xspace}

 \def\Ps      {\ensuremath{s}\xspace}

}

\def\mumu       {{\ensuremath{\Pmu^+\Pmu^-}}\xspace}

\def\W      {{\ensuremath{\PW}}\xspace}
\def\Wp     {{\ensuremath{\PW^+}}\xspace}

\def\squark    {{\ensuremath{\Ps}}\xspace}

\def\cquark    {{\ensuremath{\Pc}}\xspace}

\def\bquark    {{\ensuremath{\Pb}}\xspace}

\def\pion   {{\ensuremath{\Ppi}}\xspace}

\def\pip    {{\ensuremath{\pion^+}}\xspace}
\def\pim    {{\ensuremath{\pion^-}}\xspace}

\def\kaon    {{\ensuremath{\PK}}\xspace}
  \def\Kbar    {{\kern 0.2em\overline{\kern -0.2em \PK}{}}\xspace}

\def\Kp      {{\ensuremath{\kaon^+}}\xspace}
\def\Km      {{\ensuremath{\kaon^-}}\xspace}

  \def\Dbar    {{\kern 0.2em\overline{\kern -0.2em \PD}{}}\xspace}
\def\D       {{\ensuremath{\PD}}\xspace}

\def\Dstarzb {{\ensuremath{\Dbar^{*0}}}\xspace}

\def\Dstarm  {{\ensuremath{\D^{*-}}}\xspace}

\def\Ds      {{\ensuremath{\D^+_\squark}}\xspace}

\def\B       {{\ensuremath{\PB}}\xspace}
\def\Bbar    {{\ensuremath{\kern 0.18em\overline{\kern -0.18em \PB}{}}}\xspace}

\def\Bu      {{\ensuremath{\B^+}}\xspace}

\def\Bd      {{\ensuremath{\B^0}}\xspace}
\def\Bs      {{\ensuremath{\B^0_\squark}}\xspace}

\def\Bc      {{\ensuremath{\B_\cquark^+}}\xspace}

\def\jpsi     {{\ensuremath{{\PJ\mskip -3mu/\mskip -2mu\Ppsi\mskip 2mu}}}\xspace}
\def\psitwos  {{\ensuremath{\Ppsi{(2S)}}}\xspace}

  \def\Y#1S{\ensuremath{\PUpsilon{(#1S)}}\xspace}

\def\proton      {{\ensuremath{\Pp}}\xspace}

\def\Lbar        {{\ensuremath{\kern 0.1em\overline{\kern -0.1em\PLambda}}}\xspace}

\def\BF         {{\ensuremath{\cal B}}\xspace}

\def\BR         {\BF}
\newcommand{\decay}[2]{\ensuremath{#1\!\to #2}\xspace}         

\def\to                 {\ensuremath{\rightarrow}\xspace}

\def\AT#1     {\ensuremath{A_{\mathrm{T}}^{#1}}\xspace}           

\def\C#1      {\ensuremath{\mathcal{C}_{#1}}\xspace}                       
\def\Cp#1     {\ensuremath{\mathcal{C}_{#1}^{'}}\xspace}                    
\def\Ceff#1   {\ensuremath{\mathcal{C}_{#1}^{\mathrm{(eff)}}}\xspace}        
\def\Cpeff#1  {\ensuremath{\mathcal{C}_{#1}^{'\mathrm{(eff)}}}\xspace}       
\def\Ope#1    {\ensuremath{\mathcal{O}_{#1}}\xspace}                       
\def\Opep#1   {\ensuremath{\mathcal{O}_{#1}^{'}}\xspace}                    

\newcommand{\tev}{\ifthenelse{\boolean{inbibliography}}{\ensuremath{~T\kern -0.05em eV}\xspace}{\ensuremath{\mathrm{\,Te\kern -0.1em V}}}\xspace}
\newcommand{\gev}{\ensuremath{\mathrm{\,Ge\kern -0.1em V}}\xspace}
\newcommand{\mev}{\ensuremath{\mathrm{\,Me\kern -0.1em V}}\xspace}
\newcommand{\kev}{\ensuremath{\mathrm{\,ke\kern -0.1em V}}\xspace}
\newcommand{\ev}{\ensuremath{\mathrm{\,e\kern -0.1em V}}\xspace}
\newcommand{\gevc}{\ensuremath{{\mathrm{\,Ge\kern -0.1em V\!/}c}}\xspace}
\newcommand{\mevc}{\ensuremath{{\mathrm{\,Me\kern -0.1em V\!/}c}}\xspace}
\newcommand{\gevcc}{\ensuremath{{\mathrm{\,Ge\kern -0.1em V\!/}c^2}}\xspace}
\newcommand{\gevgevcccc}{\ensuremath{{\mathrm{\,Ge\kern -0.1em V^2\!/}c^4}}\xspace}
\newcommand{\mevcc}{\ensuremath{{\mathrm{\,Me\kern -0.1em V\!/}c^2}}\xspace}

\def\mum  {\ensuremath{{\,\upmu\rm m}}\xspace}

\def\invfb   {\ensuremath{\mbox{\,fb}^{-1}}\xspace}

\def\fs   {\ensuremath{\rm \,fs}\xspace}

\newcommand{\chisq}{\ensuremath{\chi^2}\xspace}

\newcommand{\chisqvtx}{\ensuremath{\chi^2_{\rm vtx}}\xspace}

\def\gsim{{~\raise.15em\hbox{$>$}\kern-.85em
          \lower.35em\hbox{$\sim$}~}\xspace}
\def\lsim{{~\raise.15em\hbox{$<$}\kern-.85em
          \lower.35em\hbox{$\sim$}~}\xspace}

\def\sPlot{\mbox{\em sPlot}}

\def\pt         {\mbox{$p_{\rm T}$}\xspace}

\def\bcvegpy    {\mbox{\textsc{Bcvegpy}}\xspace}

\def\evtgen     {\mbox{\textsc{EvtGen}}\xspace}

\def\geant      {\mbox{\textsc{Geant4}}\xspace}

\def\photos     {\mbox{\textsc{Photos}}\xspace}

\def\pythia     {\mbox{\textsc{Pythia}}\xspace}

\def\tell1  {TELL1\xspace}
\def\ukl1   {UKL1\xspace}

\usepackage{cite} 
\usepackage{mciteplus}

\usepackage{longtable} 

\def\FivePpi  {\ensuremath{3\pip2\pim}\xspace}

\renewcommand{\psitwos} {{\ensuremath{\Ppsi{\mathrm{(2S)}}}}\xspace}
\renewcommand{\Dbar}    {\ensuremath{\bar\D}\xspace}

\begin{document}

\renewcommand{\thefootnote}{\fnsymbol{footnote}}
\setcounter{footnote}{1}


\begin{titlepage}
\pagenumbering{roman}

\vspace*{-1.5cm}
\centerline{\large EUROPEAN ORGANIZATION FOR NUCLEAR RESEARCH (CERN)}
\vspace*{1.5cm}
\hspace*{-0.5cm}
\begin{tabular*}{\linewidth}{lc@{\extracolsep{\fill}}r}
\ifthenelse{\boolean{pdflatex}}
{\vspace*{-2.7cm}\mbox{\!\!\!\includegraphics[width=.14\textwidth]{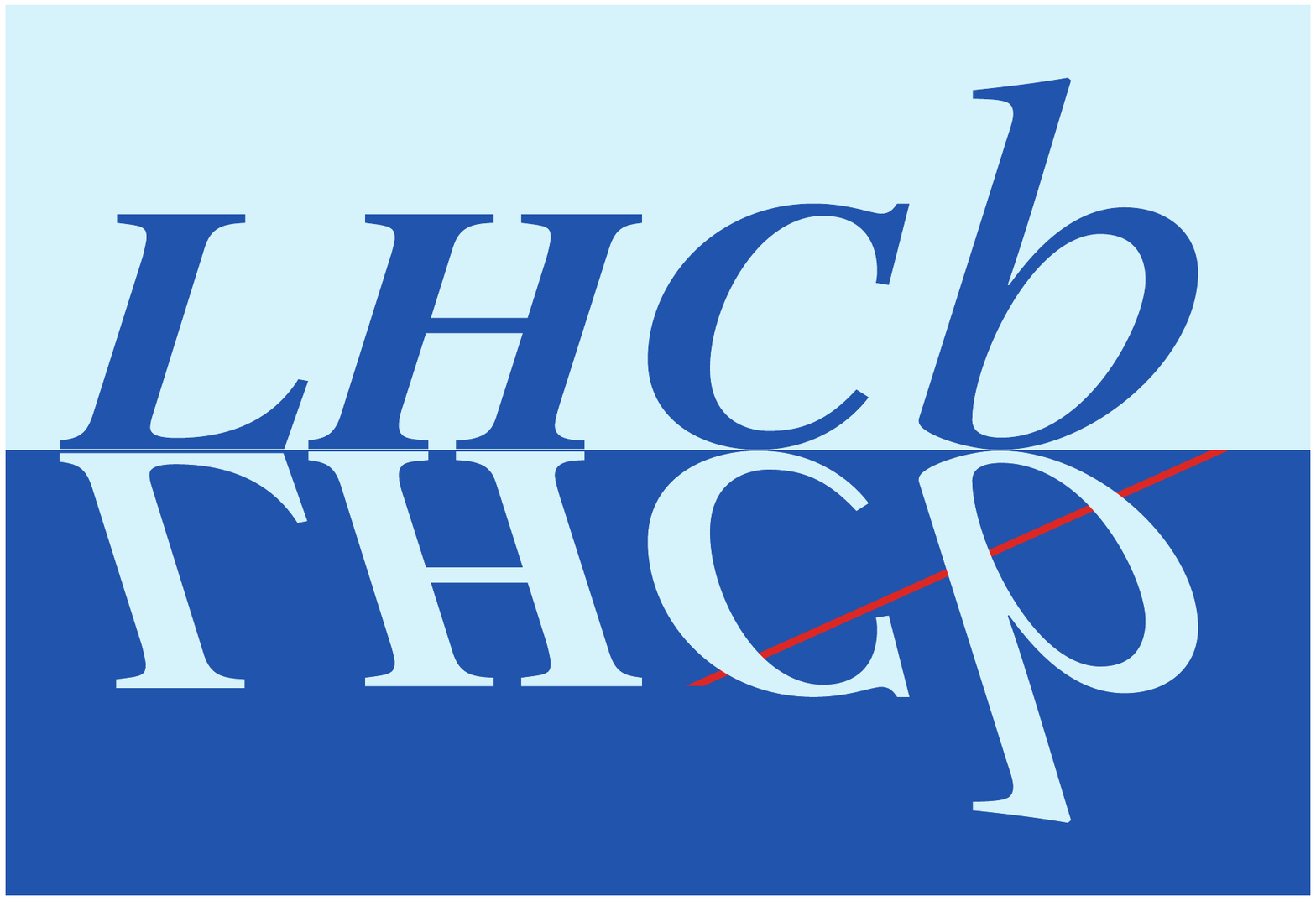}} & &}%
{\vspace*{-1.2cm}\mbox{\!\!\!\includegraphics[width=.12\textwidth]{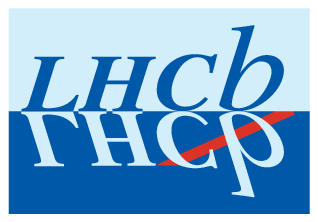}} & &}%
\\
 & & CERN-PH-EP-2014-054 \\  
 & & LHCb-PAPER-2014-009 \\  
 & &  April 1, 2014 \\ 
\end{tabular*}

\vspace*{2.0cm}

{\bf\boldmath\huge
\begin{center}
   Evidence for the decay $\Bc\to\jpsi\FivePpi$
\end{center}
}

\vspace*{1.5cm}

\begin{center}
The LHCb collaboration\footnote{Authors are listed on the following pages.}
\end{center}

\vspace{\fill}

\begin{abstract}
  \noindent
Evidence is presented for the decay $\Bc\to\jpsi\FivePpi$
using proton-proton collision data,
corresponding to an integrated luminosity  of 3\invfb,
collected with the LHCb detector.
A~signal yield of $32\pm8$ decays is found with 
a~significance of 4.5 standard deviations.
The ratio of the branching fraction of the~$\Bc\to\jpsi\FivePpi$
decay to that of the~$\Bc\to\jpsi\pip$
decay is measured to be 
\begin{equation*}
\dfrac{ \BR\left(\decay{\Bc}{\jpsi\FivePpi}\right)} 
      { \BR\left(\decay{\Bc}{\jpsi\pip}\right)} = 1.74\pm0.44\pm0.24, 
\end{equation*}
where the first uncertainty is statistical and the second is systematic.
\end{abstract}

\vspace*{2.0cm}

\begin{center}
  Submitted to JHEP
\end{center}

\vspace{\fill}

{\footnotesize 
\centerline{\copyright~CERN on behalf of the \lhcb collaboration, license \href{http://creativecommons.org/licenses/by/3.0/}{CC-BY-3.0}.}}
\vspace*{2mm}

\end{titlepage}


\newpage
\setcounter{page}{2}
\mbox{~}
\newpage

\centerline{\large\bf LHCb collaboration}
\begin{flushleft}
\small
R.~Aaij$^{41}$, 
B.~Adeva$^{37}$, 
M.~Adinolfi$^{46}$, 
A.~Affolder$^{52}$, 
Z.~Ajaltouni$^{5}$, 
J.~Albrecht$^{9}$, 
F.~Alessio$^{38}$, 
M.~Alexander$^{51}$, 
S.~Ali$^{41}$, 
G.~Alkhazov$^{30}$
% ...

{\footnotesize \it
% ...
$ ^{5}$Clermont Universit\'{e}, Universit\'{e} Blaise Pascal, CNRS/IN2P3, LPC, Clermont-Ferrand, France\\
% ...
$ ^{9}$Fakult\"{a}t Physik, Technische Universit\"{a}t Dortmund, Dortmund, Germany\\
% ...
$ ^{30}$Petersburg Nuclear Physics Institute (PNPI), Gatchina, Russia\\
% ...
$ ^{37}$Universidad de Santiago de Compostela, Santiago de Compostela, Spain\\
$ ^{38}$European Organization for Nuclear Research (CERN), Geneva, Switzerland\\
% ...
$ ^{41}$Nikhef National Institute for Subatomic Physics, Amsterdam, The Netherlands\\
% ...
$ ^{46}$H.H. Wills Physics Laboratory, University of Bristol, Bristol, United Kingdom\\
% ...
$ ^{51}$School of Physics and Astronomy, University of Glasgow, Glasgow, United Kingdom\\
$ ^{52}$Oliver Lodge Laboratory, University of Liverpool, Liverpool, United Kingdom\\
% ...
}
\end{flushleft}

\cleardoublepage


\renewcommand{\thefootnote}{\arabic{footnote}}
\setcounter{footnote}{0}



\pagestyle{plain} 
\setcounter{page}{1}
\pagenumbering{arabic}


%


\section{Introduction}
\label{sec:Introduction}

The \Bc~meson is the~only meson
consisting of two heavy quarks of different flavours.
It was discovered by the~CDF collaboration through 
the~semileptonic decay \mbox{$\decay{\Bc}{\jpsi\ell^{+}\Pnu_{\ell}\mathrm{X}}$}~\cite{CDF_Bc},
where $\mathrm{X}$~denotes possible unobserved particles.\footnote{The inclusion of charge conjugate modes is implicit throughout this paper.}
The CDF collaboration also observed the~hadronic decay mode $\Bc\to\jpsi\pip$~\cite{Aaltonen:2007gv}.
Recently, the~LHCb experiment has observed several new channels including 
\mbox{$\decay{\Bc}{\jpsi\pip\pip\pim}$}~\cite{LHCb-PAPER-2011-044},
\mbox{$\decay{\Bc}{\psitwos\pip}$}~\cite{LHCb-PAPER-2012-054},
\mbox{$\decay{\Bc}{\jpsi\Ds}$} and 
\mbox{$\decay{\Bc}{\jpsi\D_{\squark}^{*+}}$}~\cite{LHCb-PAPER-2013-010},
\mbox{$\decay{\Bc}{\jpsi\Kp}$}~\cite{LHCb-PAPER-2013-021},
\mbox{$\decay{\Bc}{\jpsi\Kp\Km\pip}$}~\cite{LHCb-PAPER-2013-047}
and 
\mbox{$\decay{\Bc}{\Bs\pip}$}~\cite{LHCb-PAPER-2013-044}.
The~lifetime of the~\Bc~meson~\cite{PDG2012,LHCb-PAPER-2013-063}
is about three times shorter than that of the~$\Bd$ and $\Bu$ mesons,
confirming the~important role played by  the~\cquark~quark in \Bc~decays.
The decays of \Bc~mesons into charmonia and light hadrons are expected to be 
well described by the~factorization approximation~\cite{Bauer:1986bm,*Wirbel:1988ft}.
In this scheme,
the~\mbox{$\decay{\Bc}{\jpsi\FivePpi}$}~decay
is characterized  by the~form factors of the~$\Bc\to\jpsi\W^{+}$
transition and the~spectral functions
for the~virtual $\W^{+}$~boson into light hadrons~\cite{Likhoded:2009ib}.
The~predictions for the~ratio of branching fractions
\begin{equation}
R_{5\Ppi}\equiv \dfrac{ \BR\left(\decay{\Bc}{\jpsi\FivePpi}\right)} 
      { \BR\left(\decay{\Bc}{\jpsi\pip}\right)}
\end{equation}  
are 0.95 and 1.1~\cite{Lesha}, using form factor calculations from Refs.~\cite{Kis1}
and~\cite{Ebert}, respectively.

In this article, the first evidence for the decay 
\mbox{$\decay{\Bc}{\jpsi\FivePpi}$} and a measurement of
$R_{5\Ppi}$~are reported.
The analysis is based on a data sample of proton-proton ($\proton\proton$)~collisions,
corresponding to an integrated luminosity of 1\invfb at a~centre-of-mass
energy of 7\tev and 2\invfb at 8\tev,
collected with the LHCb detector.

\section{Detector}
\label{sec:Detector}

The \lhcb detector~\cite{Alves:2008zz} is a single-arm forward
spectrometer covering the \mbox{pseudorapidity} range $2<\Peta <5$,
designed for the study of particles containing \bquark~or \cquark~quarks.
The detector includes a high-precision tracking system
consisting of a silicon-strip vertex detector surrounding the $\proton\proton$~interaction region,
a~large-area silicon-strip detector located
upstream of a dipole magnet with a bending power of about
$4{\rm\,Tm}$, and three stations of silicon-strip detectors and straw
drift tubes~\cite{LHCb-DP-2013-003} placed downstream.
The combined tracking system provides a momentum measurement with
relative uncertainty that varies from 0.4\% at 5\gevc to 0.6\% at 100\gevc,
and impact parameter resolution of 20\mum for
tracks with large transverse momentum.
Different types of charged hadrons are distinguished using information
from two ring-imaging Cherenkov detectors~\cite{LHCb-DP-2012-003}.
Photon, electron and hadron candidates are identified
by a~calorimeter system consisting of
scintillating-pad and preshower detectors, an electromagnetic
calorimeter and a hadronic calorimeter. 
Muons are identified by a
system composed of alternating layers of iron and multiwire
proportional chambers~\cite{LHCb-DP-2012-002}.
The trigger~\cite{LHCb-DP-2012-004} consists of a
hardware stage, based on information from the calorimeter and muon
systems, followed by a software stage, which applies a full event
reconstruction.

This analysis uses events  collected by triggers that select
the~\mumu~pair from the~\jpsi~decay with high efficiency. 
At the hardware stage either one or two muon  candidates are required to trigger the event. 
In the case of single muon triggers, the~transverse  momentum, $\pt$, 
of the~muon candidate is required to be greater than $1.5\gevc$. 
For dimuon  candidates, the product of the \pt~of muon 
candidates is required to satisfy~$\sqrt{\pt_1\pt_2}>1.3\gevc$. 
At the subsequent software trigger stage, two muons
are selected with 
an~invariant mass in the~range $2.97<m_{\mumu}<3.21\gevcc$     
and consistent with originating from a~common vertex.
The common vertex is required to be significantly
displaced from the~$\proton\proton$~collision vertices.

Simulated~$\proton\proton$~collisions are generated using 
\pythia~6.4~\cite{Sjostrand:2006za} with the 
configuration described in Ref.~\cite{LHCb-PROC-2010-056}.
Final-state QED radiative corrections are included using 
the \photos~package~\cite{Golonka:2005pn}.
The \Bc~mesons are produced by a dedicated generator, 
\bcvegpy~\cite{BCVEGPY}. The decays of all hadrons are performed 
by \evtgen~\cite{Lange:2001uf}, and a specific model is implemented 
to generate the decays $\Bc\to\jpsi\FivePpi$, assuming 
factorization~\cite{Lesha}. 
The model allows the implementation of different form factors for this decay,
calculated using QCD sum rules~\cite{Kis1} or
a~relativistic quark model~\cite{Ebert}.
These predictions lead to very similar values 
and those based on the~relativistic quark model 
are used in the~simulation. 
The~coupling of the five pion ($\FivePpi$)~system to the~virtual \Wp~is taken 
from $\Ptau^{+}$ lepton decays~\cite{Lees:2012}. 
The~interaction of the~generated particles with the~detector 
and its response are implemented using 
the~\geant~toolkit~\cite{Allison:2006ve, *Agostinelli:2002hh} as described in
Ref.~\cite{LHCb-PROC-2011-006}.

%

\section{Candidate selection}
\label{sec:EventSelection}

The~decays
\mbox{$\Bc\to\jpsi\FivePpi$}
and
\mbox{$\decay{\Bc}{\jpsi\pip}$}
are reconstructed 
using the \mbox{$\decay{\jpsi}{\mumu}$}~decay mode. 
The selection criteria chosen are similar for both channels.

All tracks are required to be 
in the pseudorapidity range $2 < \Peta < 4.9$. 
Good track quality of charged particles is ensured by requiring
the~$\chisq$ per number of degrees of freedom, $\chi^2/\mathrm{ndf}$, 
provided by the~track fit, to be less than~3. 
Suppression of fake tracks created by the~reconstruction  
is achieved by a neural network trained with simulated samples to discriminate 
between fake  tracks and tracks associated with real particles~\cite{LHCb-DP-2013-002},
ensuring the rate of fake tracks below 0.3\,\%.

Two dedicated neural networks are used for muon and pion identification. 
These~networks use the information from
the~Cherenkov detectors~\cite{LHCb-DP-2012-003}, 
muon chambers~\cite{LHCb-DP-2013-001} and
the~calorimeter system~\cite{LHCb-DP-2013-004},  
together with the tracking information.
The~momentum of the~pion candidates is required to be between 3.2\gevc and 150\gevc
in order to ensure good quality particle identification in Cherenkov detectors.
The~requirements on the neural network output are chosen
to ensure good agreement between data and simulation
and significant reduction of the~background due to misidentification.

Pairs of oppositely charged muons, originating from a~common vertex, 
are combined to form $\decay{\jpsi}{\mumu}$ candidates. 
The~\pt of each muon is required to be greater 
than 550\mevc. 
Good vertex reconstruction is ensured 
by requiring the~\chisq~of the~vertex fit, \chisqvtx, 
to be less than 20. 
To select dimuon vertices that are well-separated from
the~reconstructed $\proton\proton$~interaction vertices,
the~decay length is required to be at
least three times its uncertainty.
The~invariant mass of the~dimuon combination is required to be between 
3.020 and 3.135\gevcc. 
The~asymmetric mass range with respect to the~known \jpsi~meson mass~\cite{PDG2012}
is chosen to include the~QED radiative tail.

The selected \jpsi~candidates are combined with pions 
to form
 $\decay{\Bc}{\jpsi\FivePpi}$ and 
 $\decay{\Bc}{\jpsi\pip}$~candidates. 
The~transverse momentum of each pion 
is required to be greater than 400\mevc.
To ensure that the~pions are inconsistent with 
being directly produced in a~$\mathrm{pp}$ interaction, 
the impact parameter $\chisq$, defined as the difference between 
the $\chisq$~values of the~fits of the~$\proton\proton$~collision vertex 
formed with and without the~considered pion track,
is required to satisfy~$\chisq_{\mathrm{IP}} > 4$. When more than one primary vertex is reconstructed, 
the~vertex with the~smallest value of $\chisq_{\mathrm{IP}}$ is chosen. 
Good vertex reconstruction for the~\Bc~candidate vertex is ensured by requiring
the~$\chisqvtx/\mathrm{ndf}$ to be less than 12. 
To~suppress the~large combinatorial background in the~\mbox{$\decay{\Bc}{\jpsi\FivePpi}$} sample, 
the $\chisq$ of the~vertex fit for all 
$\jpsi\Ppi^{\pm}$~combinations, as well as for all dipion 
combinations, is required to be less than~20.
To~improve the invariant mass resolution, a~kinematic fit~\cite{Hulsbergen:2005pu} 
is performed that constrains the~$\mumu$~pair to the~known mass 
of the~$\jpsi$ meson. It is also required that the~\Bc~candidate's momentum 
vector points back to from  the~associated $\proton\proton$~interaction vertex.
When more than one $\proton\proton$~collision vertex is found,
that with the~smallest value of $\chisq_{\mathrm{IP}}$ is chosen. 
The~\chisq~per number of degrees of freedom of the fit,
$\chisq_{\mathrm{fit}}/\mathrm{ndf}$, 
is required to be less than 5. 
The~measured decay time of the~\Bc~candidate,
calculated with respect to the~associated primary vertex,
is required to be between $150\mum/c$ and 1\,mm$/c$. 

%
%
\section{Signal and normalization yields}
\label{sec:Nratio}

\begin{figure}[t]
  \setlength{\unitlength}{1mm}
  \centering
  \begin{picture}(150,120)
    \put(0,0){
      \includegraphics*[width=150mm,height=120mm%
      ]{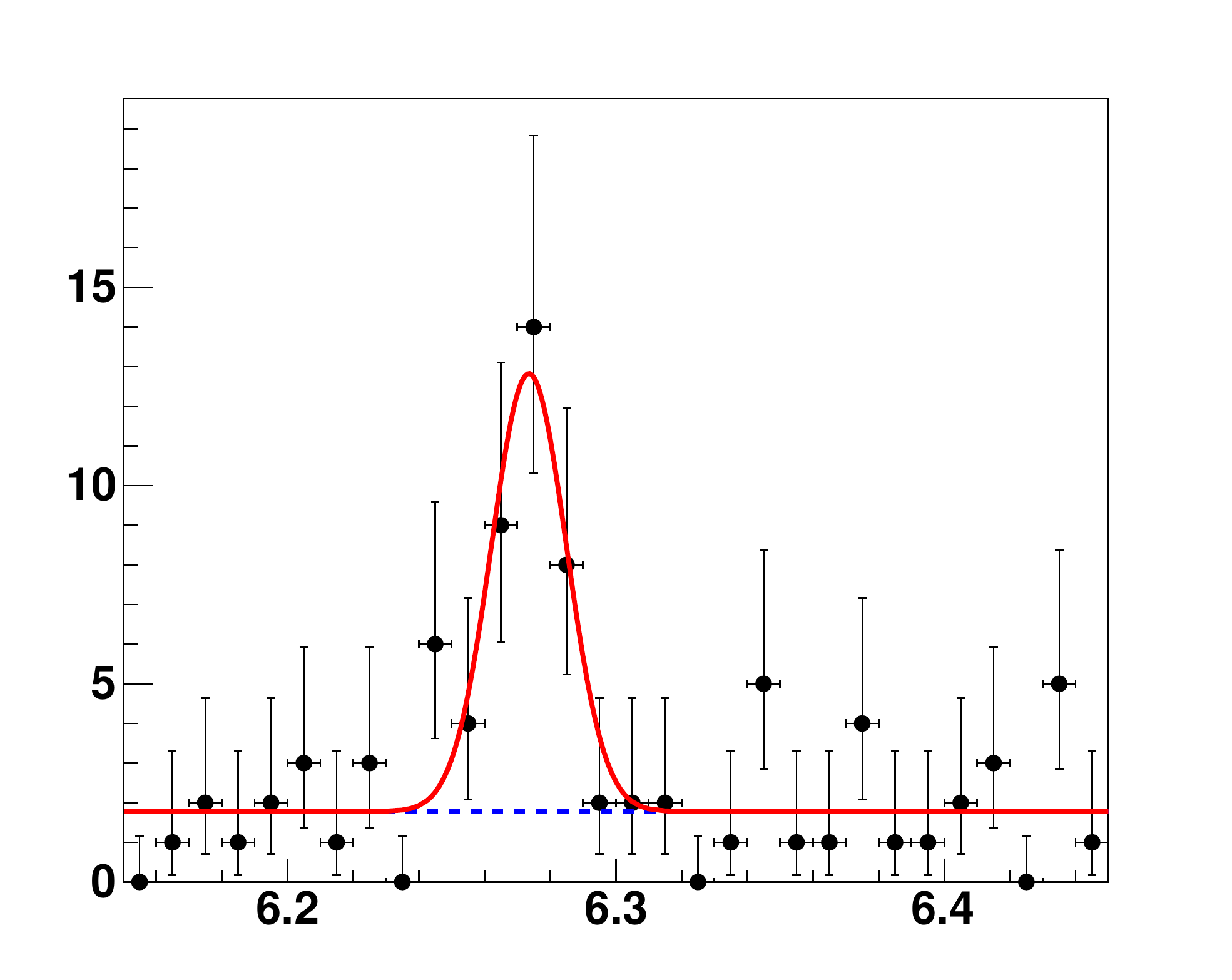}
    }
    \put( 65,0) { \large $m_{\jpsi\FivePpi}$ }
    \put(116,0) { \large $\left[ \mathrm{GeV}/c^2\right]$}
    \put( 0 , 55 ) { \large
      \begin{sideways}%
        Candidates/(10\mevcc)
      \end{sideways}%
    }
    \put(108,95){ \large LHCb }
  \end{picture}
  \caption { \small
    Mass distribution for selected
    $\decay{\Bc}{\jpsi\FivePpi}$~candidates. 
    The~result of
    a~fit using the~model described in the~text
    (red solid line)
    is shown
    together with the~background component 
    (blue dashed line).
  }
  \label{fig:Fig_1}
\end{figure}

The mass distribution for the selected $\jpsi\FivePpi$ candidates
is shown in Fig.~\ref{fig:Fig_1}. 
To~estimate the signal yield, an~extended maximum likelihood fit 
to the unbinned mass distribution is made.
The $\decay{\Bc}{\jpsi\FivePpi}$~signal is modelled by a~Gaussian distribution 
and the~background by a~constant function. 
The~fit results for the~fitted mass and mass resolution
of \Bc~signal,
$m_{\Bc} $ and
$\sigma_{\Bc}$, and signal yield $N_{\Bc\to\jpsi\FivePpi}$,
are listed in Table~\ref{tab:signal_fitres_5pi},

The statistical significance for the observed signal is determined as
\mbox{$\mathcal{S}_{\upsigma}=\sqrt{-2\log{\frac{\mathcal{L}_\mathrm{B}}{\mathcal{L}_{\mathrm{S+B}}}}}$}
where ${\mathcal{L}_{\mathrm{S+B}}}$ and ${\mathcal{L}_{\mathrm{B}}}$ denote the
likelihood associated with the signal-plus-background and background-only hypothesis,
respectively.
The likelihoods are calculated with the~peak position fixed to
the~known mass
of \Bc~meson~\cite{PDG2012,LHCb-PAPER-2013-010}
and the~mass resolution
fixed to~10.1\mevcc as expected from simulation.
The statistical significance of the~\mbox{$\mathrm{B_c^+}\to\jpsi\FivePpi$} 
signal is 4.5~standard deviations.

\begin{figure}[t]
  \setlength{\unitlength}{1mm}
  \centering
  \begin{picture}(150,120)
    \put(0,0){
      \includegraphics*[width=150mm,height=120mm%
      ]{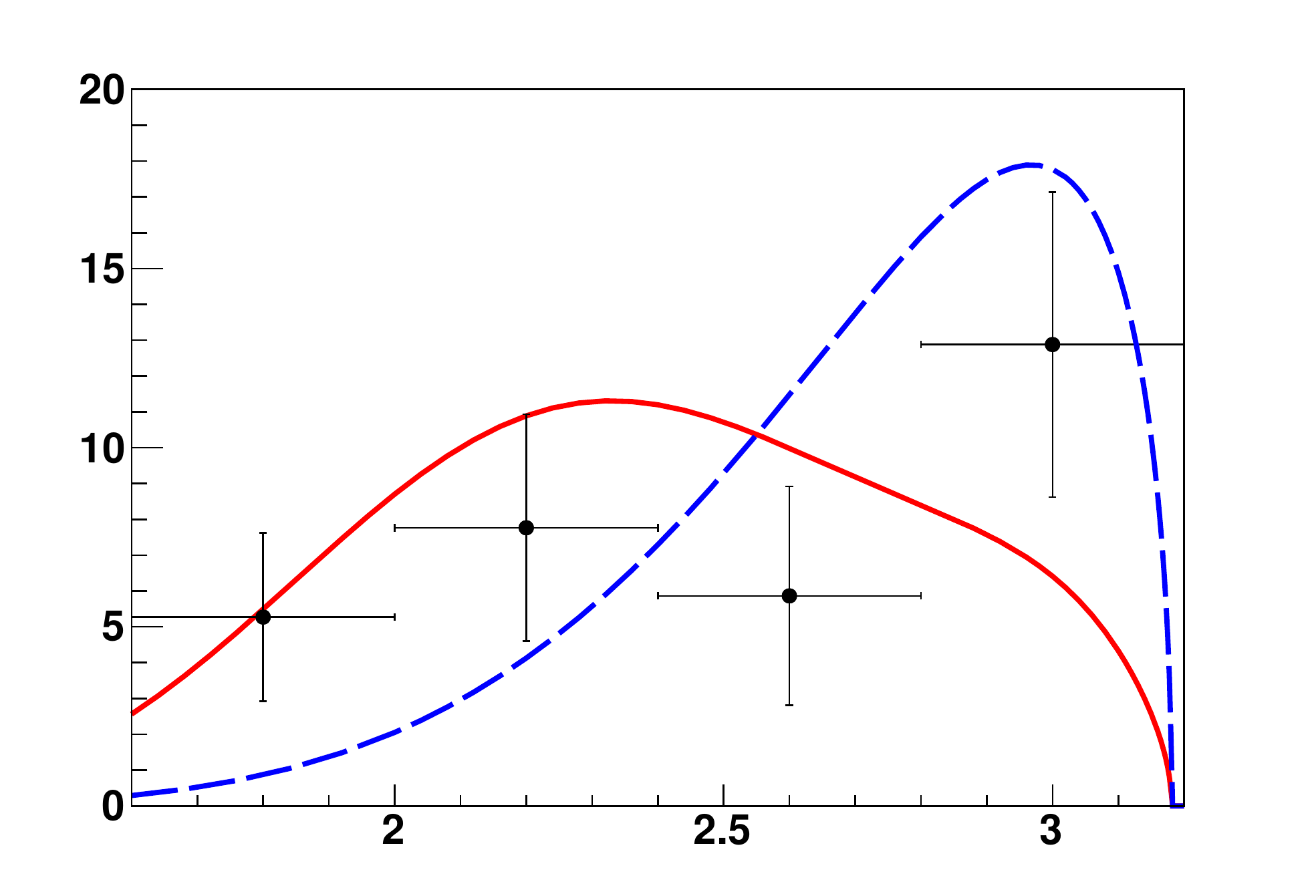}
    }
    \put( 70,0 ) { \large $m_{\FivePpi}$}
    \put(120,0 ) { \large $\left[ \mathrm{GeV}/c^2\right]$}
    \put(  -2,65) { \large
      \begin{sideways}%
        Yield/(400\mevcc)
      \end{sideways}%
    }
    \put(20,95 ){ \large LHCb }
  \end{picture}
  \caption { 
    \small
    Background-subtracted distribution of five-pion mass
    from $\decay{\Bc}{\jpsi\FivePpi}$ events (points with error bars).
    The model prediction from Ref.~\cite{Lesha} is shown 
    by a~red solid line,
    and the expectation
    from the~phase space model is shown by a~blue dashed line.
   }
  \label{fig:Fig_2}
\end{figure}
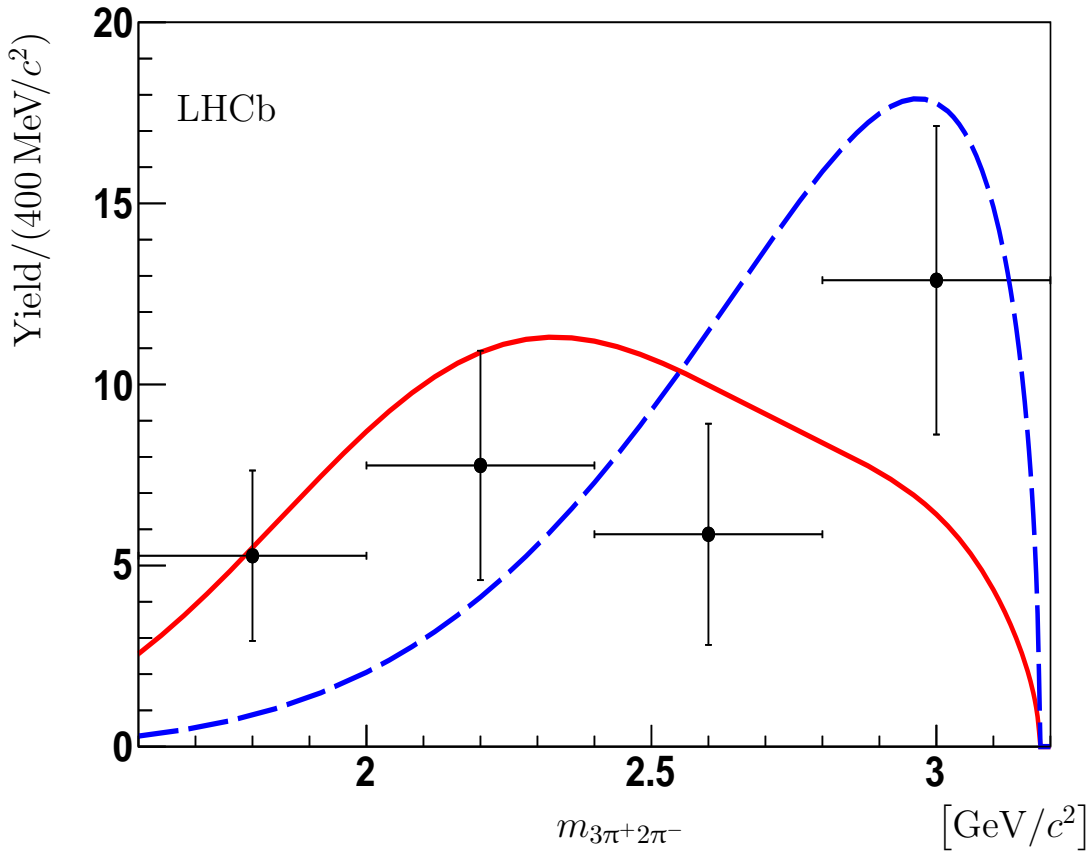

For the selected \Bc~candidates, the existence of resonant structures is searched for
in the~$\pip\pim$, $\pip\pip\pim$, $\pip\pim\pim$, $2\pip2\pim$, 
$\FivePpi$ and $\jpsi\pip\pim$~combinations of final state particles  
using the \sPlot~technique~\cite{Pivk:2004ty},
with the~reconstructed $\jpsi\FivePpi$~mass
as discriminating variable,
to subtract the~background.
No~significant narrow structures are observed; in particular, 
no indication of a~contribution from \mbox{$\decay{\Bc}{\psitwos\pip\pip\pim}$}, 
followed by the \mbox{$\decay{\psitwos}{\jpsi\pip\pim}$} decay, is seen. 
The background-subtracted five-pion mass distribution is shown in Fig.~\ref{fig:Fig_2},
along with the~theoretical prediction in Ref.~\cite{Lesha}, which describes the data well.
The~consistency between data 
and the~model prediction
is estimated using a~\chisq-test and gives a~$p$-value of~14\,\%.
The corresponding $p$-value for the~phase space decay model is 4\,\%.

\begin{table}[t]
\centering
\caption{\small
  Signal parameters of the unbinned
  extended maximum likelihood
  fit to the~$\jpsi\FivePpi$~mass distribution. Uncertainties are statistical only.}
\vspace*{3mm}
\begin{tabular*}{0.75\textwidth}{@{\hspace{15mm}}lc@{\extracolsep{\fill}}c@{\hspace{15mm}}}
  \multicolumn{2}{c}{Parameter}  & Value \\ \hline
  $m_{\Bc} $                        &  $\left[\mevcc\right]$ & $6273 \pm 3\phantom{000}$   \\
  $\sigma_{\Bc}                   $ &  $\left[\mevcc\right]$ & $11.4     \pm 3.4\phantom{0}$     \\
  $N_{\Bc\to\jpsi\FivePpi}                     $ &                        & $32    \pm 8\phantom{0}$     \\
\end{tabular*}
\label{tab:signal_fitres_5pi}
\end{table}

\begin{figure}[t]
  \setlength{\unitlength}{1mm}
  \centering
  \begin{picture}(150,120)
    \put(0,0){
      \includegraphics*[width=150mm,height=120mm%
      ]{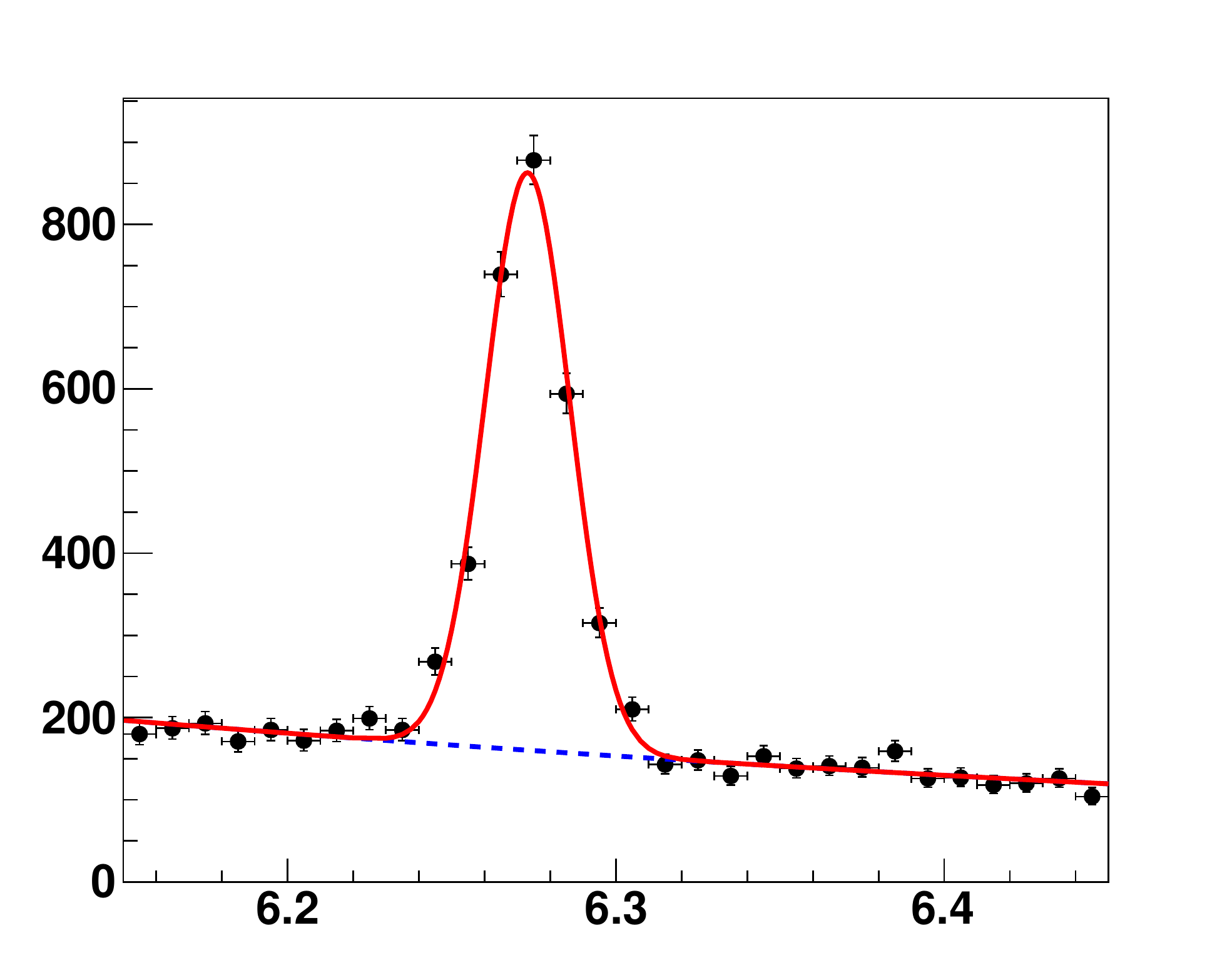}
    }
    \put( 65,0) { \large $m_{\jpsi\pip}$ }
    \put(116,0) { \large $\left[ \mathrm{GeV}/c^2\right]$}
    \put( 0 , 55 ) { \large
      \begin{sideways}%
        Candidates/(10\mevcc)
      \end{sideways}%
    }
    \put(108,95){ \large LHCb }
  \end{picture}
  \caption { \small
    Mass distribution for selected
    $\decay{\Bc}{\jpsi\pip}$~candidates. 
    The~result of
    a~fit using the~model described in the~text
    (red solid line)
    is shown
    together with the~background component 
    (blue dashed line).
  }
  \label{fig:Fig_3}
\end{figure}

The mass distribution of the selected  $\Bc\to\jpsi\pip$~candidates is shown in Fig.~\ref{fig:Fig_3},
together with the result of an~extended unbinned maximum likelihood fit. 
The \Bc~signal is modelled by a~Gaussian distribution and the background
by an exponential function. The fit gives a yield of $2271\pm63$ events.

%

\section{Efficiency and systematic uncertainties}
\label{sec:syst}

The overall efficiency for each decay is the product of
the~geometrical acceptance of the~detector,
reconstruction, selection and trigger
efficiencies. These are estimated using simulation
and the~ratio of the efficiencies is found to be
\begin{equation}
  \dfrac{\Pvarepsilon(\Bc\to\jpsi\pip)}{\Pvarepsilon(\Bc\to\jpsi\FivePpi)} = 123.8 \pm 5.6,
  \label{eq:eff}
\end{equation}
where the uncertainty reflects the size of the simulated sample.
The large difference in
efficiencies is due to the~reconstruction of four additional low-\pt~pions in the~$\decay{\Bc}{\jpsi\FivePpi}$~mode.
The efficiencies for the data samples collected at
a~centre-of-mass energy of 7\tev and 8\tev are
found to be similar and
a~luminosity-weighted average is used,
with the~corresponding systematic
uncertainty discussed below.

Many sources of systematic uncertainty cancel in the ratio, in particular those 
related to the muon and $\jpsi$ reconstruction and identification.
Those that do not cancel are discussed below
and summarized in Table~\ref{table:syst}.

\begin{table}[t]
  \centering
  \caption{ 
    \small
    Relative systematic uncertainties
    for the~ratio $R_{5\Ppi}$. 
    The~total uncertainty is the~quadratic 
    sum of the~individual components.
  } \label{table:syst}
  \vspace*{3mm}
  \begin{tabular*}{0.8\textwidth}{@{\hspace{10mm}}l@{\extracolsep{\fill}}c@{\hspace{10mm}}}
    Source & Uncertainty~$\left[\%\right]$
    \\
    \hline
    Fit model                              &  6.6  \\
    Decay model                            &       \\ 
    ~~$m_{3\pip2\pim}$~reweighting            &  7.7  \\ 
    ~~$\psitwos$ mass veto                 &  3.1  \\
    Data-simulation agreement     &  \\ 
    ~~Hadron interactions                  &  $4\times2.0$  \\
    ~~Track quality selection              &  $4\times0.6$  \\
    ~~Trigger                              &  1.1           \\ 
    ~~Pion identification                  &  0.7           \\ 
    ~~Selection variables                  &  1.0            \\
    $\Bc$ lifetime                    & 0.9 \\
    Stability for various data taking conditions & 2.5 \\ 
    Acceptance                        &  0.8 \\ 
    \hline
    Total    &  13.9\phantom{0}
  \end{tabular*}
 \end{table}

A~systematic uncertainty arises from the~imperfect knowledge of the~shape 
of the~signal and background in the~$\jpsi\FivePpi$ and $\jpsi\pip$~mass distributions. 
The dependence of the~signal yields on the fit model
is studied 
by varying the signal and background parameterizations.
This is assessed by using Crystal Ball~\cite{Skwarnicki:1986xj} 
and double-sided Crystal Ball~\cite{LHCb-PAPER-2011-013}  
functions for 
the~parameterization of the \Bc~signals. 
The~background parametrization is performed using both exponential and polynomial functions.      
The~maximum observed change of 6.6\,\%
in the ratio of 
\mbox{$\decay{\Bc}{\jpsi\FivePpi}$} and 
\mbox{$\decay{\Bc}{\jpsi\pip}$}~yields 
is assigned  as a~systematic uncertainty. 

To assess the systematic uncertainty related to  the \mbox{$\decay{\Bc}{\jpsi\FivePpi}$}~decay model 
used in the simulation~\cite{Lesha}, the reconstructed mass distribution of the five-pion system 
in simulated events is reweighted to reproduce the distribution observed in data.
As a~cross-check the~efficiency is also recalculated using
a~phase space model for the~\mbox{$\decay{\Bc}{\jpsi\FivePpi}$}~decays.
There is a~maximal change in efficiency of 7.7\,\%, which is taken as 
the~systematic 
uncertainty for the~decay model. In~addition, the~analysis is 
repeated with the~removal of all \Bc~candidates 
where the~$\jpsi\pip\pim$ mass is compatible with
originating from $\decay{\psitwos}{\jpsi\pip\pim}$~decays. 
The~observed difference of 
3.1\,\% is assigned as an~additional systematic uncertainty.
    
A large class of uncertainties arises from the differences between data and simulation,
in particular those affecting the efficiency for reconstruction of charged-particle tracks.
The largest of these arises from the simulation of hadronic interactions in the detector, 
which has an~uncertainty  of $2\,\%$ per track~\cite{LHCb-DP-2013-002,LHCb-PUB-2011-025,LHCb-PAPER-2010-001}. 
An~additional uncertainty associated with the~track quality requirements for the~additional
four pions in the~signal decay is estimated to be $0.6\,\%$ per track~\cite{LHCb-PAPER-2013-010,LHCb-PAPER-2013-047}. 
The trigger efficiency for events with $\decay{\jpsi}{\mumu}$
produced in beauty hadron decays is studied on data in high-yield 
modes~\cite{LHCb-PAPER-2012-010,LHCb-PAPER-2013-010}
and a~systematic uncertainty of 1.1\,\% is assigned based on 
the~comparison of the~ratio of trigger  efficiencies
for  high-yield samples of $\Bu\to\jpsi\Kp$ and 
$\Bu\to\psitwos\Kp$~decays on data and simulation~\cite{LHCb-PAPER-2012-010}.

The systematic uncertainty associated with pion identification 
is studied using a~sample of $\decay{\Bu}{\jpsi\Kp\pip\pim}$~decays.
The efficiency to identify a $\pip\pim$~pair is compared for data and simulation. 
This comparison shows a 0.35\% difference between the~data and simulation 
in the efficiency to identify a~pion pair. 
As a~result of this study an~uncertainty of 0.7\,\% is assigned for the~four additional pions in the analysis.

The transverse momentum and rapidity spectra for
  the~selected $\decay{\Bc}{\jpsi\pip}$~candidates,
  as well their daughter \jpsi~mesons  and pions, are found to be
  in good agreement with the~predictions from the~\bcvegpy~generator.
Good agreement in efficiencies determined from the~data and simulation has been observed for all
variables  used in the~selection of $\decay{\Bc}{\jpsi\pip}$~candidates.
The~differences  do not exceed 1\,\%,
which is used as a~conservative estimate for 
the~systematic uncertainty from the selection variables.  
The agreement between data and simulation 
has also been cross-checked using the~$\decay{\Bc}{\jpsi\FivePpi}$~signal
by varying the selection criteria to the values that correspond to 
a~$20\,\%$ change in the signal yield in simulation. 
No unexpectedly large deviation is found.

The different acceptance as a~function of decay time for 
the~$\decay{\Bc}{\jpsi\FivePpi}$ and  
$\decay{\Bc}{\jpsi\pip}$~decay modes results
in an~additional systematic uncertainty related 
to the~imprecise knowledge of the \Bc~lifetime.
To~assess the related uncertainty,   the decay time distributions for 
simulated events are reweighted after changing 
the \Bc~lifetime by one standard  deviation around the~value of 
$509\pm8\pm12\fs$~\cite{LHCb-PAPER-2013-063} measured by LHCb and 
the~efficiencies are recomputed.  
The~observed 0.9\,\%~variation in 
the ratio of efficiencies is used as the~systematic uncertainty.

The uncertainty related to the stability of the analysis results against
variations of the~detector and trigger configurations occuring in different
data-taking periods are tested by studying the ratio of the yields of  
$\decay{\Bu}{\jpsi\Kp\pip\pim}$ and $\decay{\Bu}{\jpsi\Kp}$ 
decays as a~function of the~data-taking period. 
According to this study an additional systematic 
uncertainty of 2.5\,\%~is assigned~\cite{LHCb-PAPER-2013-010}.

The last systematic uncertainty originates from the dependence 
of the~geometrical acceptance on  both the beam crossing angle and the position 
of the luminosity region. The~resulting 0.8\,\%~difference in the efficiency 
ratios is taken as an estimate of the~systematic uncertainty.

A~summary of systematic uncertainties is presented in Table~\ref{table:syst}.
The~total systematic uncertainty on the ratio of the branching fractions
$R_{5\Ppi}$ is $13.9\,\%$.


\section{Results and summary }
\label{sec:result}
The first evidence for the decay $\Bc\to\jpsi\FivePpi$ is found
using $\proton\proton$~collisions, 
corresponding to an~integrated luminosity of 3\invfb, collected with the~LHCb detector
A~signal yield of $32 \pm 8$~events is found. 
The~significance, taking into account the~systematic uncertainties
due to the~fit function, peak position and mass resolution
in the~fit, is estimated to be 4.5~standard deviations.

Using the $\Bc\to\jpsi\pip$ mode as a~normalization channel,
the ratio of branching fractions is calculated as
\begin{equation}
  R_{5\Ppi} = 
\dfrac{N\left(\Bc\to\jpsi\FivePpi\right)}
      {N\left(\Bc\to\jpsi\pip\right)}
\times
\dfrac{\Pvarepsilon(\Bc\to\jpsi\pip)}{\Pvarepsilon(\Bc\to\jpsi\FivePpi)},
\end{equation}
where $N$~is the number of reconstructed decays
obtained from the fit described in Sect.~\ref{sec:Nratio}
and the~efficiency ratio is taken from Eq.~\eqref{eq:eff}.
The ratio of branching fractions is measured to be
\begin{equation*}
\dfrac{\BR\left(\Bc\to\jpsi\FivePpi\right)}
      {\BR\left(\Bc\to\jpsi\pip\right)} =1.74\pm0.44\pm0.24,
\end{equation*}
where the first uncertainty is statistical and the second is systematic. 
The result is in agreement with theoretical predictions~\cite{Lesha} of 0.95 and 1.1 
using the form factors  from Refs.~\cite{Kis1} 
and~\cite{Ebert}, respectively.
This result is also 
consistent with analogous measurements in \Bd and \Bu~meson decays~\cite{PDG2012}
\begin{eqnarray*}
\dfrac {\BR\left(\decay{\Bd}{\Dstarm  3\pip 2\pim}\right)}
       {\BR\left(\decay{\Bd}{\Dstarm   \pip      }\right)} & = & 1.70\pm0.34, \\   
\dfrac {\BR\left(\decay{\Bu}{\Dstarzb 3\pip 2\pim}\right)}
       {\BR\left(\decay{\Bu}{\Dstarzb  \pip      }\right)} & = & 1.10\pm0.24,
\end{eqnarray*}
as expected from factorization.

\section*{Acknowledgements}

We thank A.K.~Likhoded and A.V.~Luchinky for fruitful
discussions about the dynamics of \Bc~decays.
We express our gratitude to our colleagues in the CERN
accelerator departments for the excellent performance of the LHC. We
thank the technical and administrative staff at the LHCb
institutes. We acknowledge support from CERN and from the national
agencies: CAPES, CNPq, FAPERJ and FINEP (Brazil); NSFC (China);
CNRS/IN2P3 and Region Auvergne (France); BMBF, DFG, HGF and MPG
(Germany); SFI (Ireland); INFN (Italy); FOM and NWO (The Netherlands);
SCSR (Poland); MEN/IFA (Romania); MinES, Rosatom, RFBR and NRC
``Kurchatov Institute'' (Russia); MinECo, XuntaGal and GENCAT (Spain);
SNSF and SER (Switzerland); NASU (Ukraine); STFC and the Royal Society (United
Kingdom); NSF (USA). We also acknowledge the support received from EPLANET, 
Marie Curie Actions and the ERC under FP7. 
The Tier1 computing centres are supported by IN2P3 (France), KIT and BMBF (Germany),
INFN (Italy), NWO and SURF (The Netherlands), PIC (Spain), GridPP (United Kingdom).
We are indebted to the communities behind the multiple open source software packages on which we depend.
We are also thankful for the computing resources and the access to software R\&D tools provided by Yandex LLC (Russia).

\addcontentsline{toc}{section}{References}
\setboolean{inbibliography}{true}
\bibliographystyle{LHCb}
\bibliography{main,LHCb-PAPER,LHCb-CONF,LHCb-DP,local}


\end{document}